\documentclass[reqno]{article}
\usepackage{graphicx}
\usepackage{amsmath,amsthm,amssymb}
\chardef\bslash=`\\

\newtheorem{thm}{Theorem}[section]

\newtheorem{lem}[thm]{Lemma}
\newtheorem{prop}[thm]{Proposition}

\theoremstyle{definition}
\newtheorem{defn}{Definition}[section]
\theoremstyle{remark}
\newtheorem*{rem}{Remark}

\numberwithin{equation}{section}

\newcommand{\eval}[2][\right]{\relax\ifx#1\right\relax \left.\fi#2#1\rvert}

\let\norm=\enVert

\begin{document}
\title{On existence of mini-boson stars}
\author{Piotr Bizo\'n\thanks{Institute of Physics, Jagellonian University,
Krak\'ow, Poland}\, and Arthur Wasserman\thanks
{Department of Mathematics, University of Michigan, Ann Arbor, Michigan}}
\maketitle
\begin{abstract}
We prove the existence of a countable family of globally regular
solutions of spherically symmetric Einstein-Klein-Gordon equations.
These solutions, known as mini-boson stars, were discovered
numerically many years ago.
\end{abstract}
\section{Introduction}
\label{intro}
Boson stars are compact gravitationally bound
soliton-like equilibrium configurations of bosonic fields. The
simplest kind of boson star, which is made up of a self-gravitating
free complex massive scalar field, was conceived over thirty years
ago by Kaup~\cite{kaup} and Ruffini and Bonazzola~\cite{rb} who found
numerically the ground state solution to the spherically symmetric
Einstein-Klein-Gordon (EKG) equations. A decade later the systematic
numerical analysis of these equations was performed by Friedberg,
Lee, and Pang~\cite{flp} who rediscovered and extended the results
of~\cite{kaup, rb}, in particular they found a countable sequence of
excited states.

The aim of this paper is to give a rigorous proof of existence of
solutions found in~\cite{kaup,rb,flp}. In the physics literature
these solutions are usually referred to
as mini-boson stars ('mini' because  they are tiny
objects with mass $\sim$ $1\over Gm$, where $m$ is the boson mass).
What are the motivations for studying such objects? Let us mention
three possible reasons varying from physical to purely mathematical.
First, most theories of elementary particles
predict the existence of massive bosons which interact weakly with
baryonic matter. To the extent one believes in these models, one
should accept their consequences, like boson stars. From this
standpoint, the recent surge of interest in boson stars is largely
due to the suggestion that the dark matter could be bosonic since
then some fraction of the missing mass of the universe would float
around in the form of boson stars. Second, even if massive scalar
fields do not exist in nature, they provide one of the simplest
fundamental matter sources for the Einstein equations and, as such,
are ideal theoretical ``laboratories'' for studying the dynamics of
gravitational collapse. Mathematically, these studies amount to the
analysis of the Cauchy problem for the EKG equations. Boson stars
play an important role in this context as candidates for intermediate
or final attractors of dynamical evolution. Finally, and admittedly
most
interestingly for us, mini-boson stars are {\em non-perturbative}
solutions of the EKG equations in the sense that they have no regular
flat-spacetime limit (one
manifestation of this property is the fact mentioned above that the
total  mass of a mini-boson star is inversely proportional to the
gravitational constant $G$). In this respect mini-boson stars are
similar to the Bartnik-McKinnon solutions of the Einstein-Yang-Mills
equations~\cite{bm}. However, in contrast to the Bartnik-McKinnon
solutions, the mini-boson stars are not static: although the metric
and the stress-energy tensor of the scalar field are
time-independent, the scalar field iself has the form of a standing
wave $\phi(r,t)=e^{i \omega t}\tilde\phi(r)$. This fact has an important
consequence at the ode level, namely the lapse function does not
decouple from the Klein-Gordon equation and the hamiltonian
constraint which means that we have to deal with a 4-dimensional
(nonautonomous) dynamical system\footnote{For
comparison, the static spherically
symmetric Einstein-Yang-Mills equations reduce (within the purely
magnetic ansatz) to
a 3-dimensional (nonautonomous) dynamical system.}.
Below we analyze this system using a shooting method which is similar
in spirit (but quite different in implementation) to
the proof of
existence of the Bartnik-McKinnon solutions~\cite{sw}.

The paper is organized as follows. In Section~2 we derive the
field equations together with the boundary conditions  and discuss
some basic properties of solutions. We also formulate the main
theorem and sketch the heuristic idea of its proof. In Section~3
we prove the local existence of solutions near the origin. In
Sections~4 and~5 we discuss the limiting behavior of solutions for
small and large values of the shooting parameter, respectively. In
Section~6 we derive the asymptotics of globally regular solutions.
Section~7 contains some technical results concerning the behavior
of singular solutions. Finally, in Section~8, using the results of
Sections~4-7, we complete the proof of the main theorem by a
shooting argument.
\section{Preliminaries}
The action for the EKG system is given by
\begin{equation}
S = \int d^4 x \sqrt{-g}\left(\frac{R}{16 \pi G} - \frac{1}{2}
\partial_a\phi^* \partial^a\phi - \frac{1}{2} m^2\phi^*\phi \right),
\end{equation}
where $R$ is the scalar curvature of the spacetime metric $g_{ab}$,
$\phi$ is the complex scalar field, and $m$ is a real constant  called the
boson mass. The associated field equations are the Einstein equations
\begin{equation}
R_{ab}-\frac{1}{2} g_{ab} R = 8 \pi G T_{ab}
\end{equation}
with the stress-energy tensor of the scalar field
\begin{equation}
T_{ab}=\frac{1}{2} (\partial_a \phi^*\partial_b\phi+
\partial_a \phi\,\partial_b\phi^*)-\frac{1}{2} g_{ab}
(g^{cd}\partial_c\phi^*\partial_d\phi + m^2 \phi^*\phi ),
\end{equation}
and the Klein-Gordon equation
\begin{equation}
(\square- m^2) \phi = 0,
\end{equation}
where $\square$ is the d'Alembertian operator associated with the
metric $g_{ab}$. Now, we assume that the fields are spherically
symmetric. We write the metric using areal radial coordinate and
polar slicing
\begin{equation}
ds^2 = -e^{-2\delta} A dt^2 + A^{-1} dr^2 + r^2 d \Omega^2,
\label{METRIC}
\end{equation}
where $d\Omega^2$ is the standard metric on the unit two-sphere, and
$A$ and $\delta$ are functions of $(t,r)$. In this parametrization
the (relevent components of) Einstein equations have the particularly
simple form
\begin{eqnarray}
\partial_r A &=& \frac{1-A}{r} - 8 \pi G r\,T_{00},\\
{\partial_r\delta} &=& -4 \pi G r A^{-1} (T_{00}+T_{11}),\\
\partial_t A &=& -8 \pi G r e^{-\delta} A T_{01},
\end{eqnarray}
where the components of stress-energy tensor $T_{ab}$ are expressed
in the orthonormal frame determined by the metric (2.5)
($e_0=e^{\delta} A^{-1/2} \partial_t$, $e_1=A^{1/2}\partial_r$).
From (2.3) we obtain
\begin{eqnarray}
T_{00} &=& \frac{1}{2} (A |\partial_r\phi|^2+A^{-1}
e^{2\delta} |\partial_t\phi|^2 + m^2 |\phi|^2),\\
T_{11} &=& \frac{1}{2} (A |\partial_r\phi|^2+A^{-1} e^{2\delta}
|\partial_t\phi|^2 -m^2 |\phi|^2),\\
T_{01} &=& \frac{1}{2} e^{\delta} (\partial_r\phi^*
\partial_t\phi+\partial_r\phi\, \partial_t\phi^*).
\end{eqnarray}
The remaining components of Einstein's equations are equivalent to
the Klein-Gordon equation.

For the scalar field $\phi$ we  assume the standing wave ansatz
$\phi(r,t)=\exp(i \omega t)\tilde \phi(r)$,
where $\omega$ is a real constant. Then, due to the
$U(1)$ symmetry of the action, the stress-energy tensor and the
metric are time-independent. Morever, $T_{01}=0$ so Eq.~(2.8) is
trivially satisfied. In terms of the dimensionless variables
\begin{equation}
x=m r, \qquad f(x)= \sqrt{4 \pi G} \,\tilde \phi(r)
\end{equation}
and the auxiliary variable
\begin{equation}
C(x)=\frac{\omega}{m} A^{-1} e^{\delta},
\end{equation}
Eqs.~(2.4),(2.6) and (2.7) reduce to the following system of ordinary
differential equations (hereafter prime denotes $\frac{d}{dx}$)
\begin{subequations}
\begin{eqnarray}
\left(\frac{x^2 f'}{C}\right)' &=& \frac{x^2}{AC} (1-AC^2) f,\\
A' &=&
\frac{1-A}{x}-x ( A {f'}^2+A C^2 f^2 + f^2),\\
C' &=& \frac{C}{x A}(A
-1+ x^2 f^2).
\end{eqnarray}
\end{subequations}
Instead of $A$, it is sometimes convenient to use the ``mass''
function $M(x)$ defined by $A(x)=1-2 M(x)/x$. From (2.14b) we have
\begin{equation}
M(x)=\frac{1}{2}\int_0^x s^2 ( A{f'}^2+AC^2f^2+f^2) ds.
\end{equation}
A spacetime is said to be asymptotically flat if $\delta(\infty)$ is
finite and
\begin{equation}
\lim_{x \rightarrow \infty}M(x)=M_{\infty} < \infty.
\end{equation}
The limiting value $M_{\infty}$ is interpreted as the total mass of a
solution (in our case it is measured in units $\frac{1}{Gm}$). In
Section~6 we will show that the finiteness of mass implies that $C$
has a finite limit and $f$ decays exponentially as $x \rightarrow\infty$.

Besides the singularity at infinity, the field equations (2.14) have
the fixed singular point at $x=0$
and a moving singularity at $\bar x$, where $A(\bar x)=0$.
Regularity of solutions at $x=0$  requires the following behavior
\begin{equation}
f(x)=a+O(x^2), \quad A(x)=1+O(x^2), \quad C(x)=\alpha+O(x^2),
\end{equation}
where $a=f(0)$ and $\alpha=C(0)$ are arbitrary parameters (assumed
positive without loss of generality). In Sect.~3 we will show that
these parameters determine uniquely a smooth local solution to
Eqs.(2.14).
\begin{defn}
The solution of Eqs.(2.14) starting at $x=0$ with the behavior (2.17)
is called the $\mathbf{a\!-\!orbit}$.
\end{defn}
In the following whenever we write 'a solution' we always mean the
$a$-orbit. Also when we write that some property holds for all $x$ we
always mean for all $x \geq 0$. We will frequently refer to the
behavior of $a$-orbits in the $(f,f')$-plane; when we write, say,
that
the $a$-orbit enters the first quadrant (Q1 for brevity), we mean
that the projection of the $a$-orbit in the $(f,f')$-plane does so.
\begin{defn}
The $a$-orbits which exist for all $x$ and are asymptotically flat
are called {\bf
globally regular}.
\end{defn}

Now, we are ready to formulate our main result:
\begin{thm}
For each $\alpha>1$, there is a decreasing sequence of parameters
$a_n$ ($n=0, 1, 2,...$) such that the corresponding $a_n$-orbits are
globally regular. The index $n$ labels the number of nodes of the
function $f(x)$.
\end{thm}
This theorem makes rigorous the numerical
results obtained in~\cite{kaup,rb,flp}.
Notice that although the $a$-orbits are determined by two
parameters, only the parameter $a$  has to be fine-tuned so the shooting
is essentially one-dimensional.

In order to prepare the ground for the proof of Theorem~2.1 we
discuss
now some
elementary global properties of $a$-orbits.
\begin{lem}
$A(x)<1$ for all $x>0$ unless $f(x)\equiv 0$.
\end{lem}
\begin{proof} From (2.14b), $A'(x_0)<0$ if $A(x_0)=1$ so $A$ cannot
cross $1$ from below. Since $A(x)<1$ for small $x$, the lemma
follows.
\end{proof}
\begin{lem}
An $a$-orbit exists as long as $A(x)>0$.
\end{lem}
\begin{proof}
If $A(x)>0$ for $x<\bar x <\infty$, then  $\lim_{x \rightarrow \bar x}M(x)$
exists (because $M'>0$ and $M(x)<x<\bar x$) so
$\lim_{x\rightarrow \bar x}A$ exists as well. We will show that the orbit can
be continued beyond $\bar x$ provided that
$\lim_{x \rightarrow \bar x} A(x)>0$.
Since $0<A<1$, the only obstruction to extending the
solution is the possibility that $C$,$f$, or $f'$ might be unbounded.
To see that $f$ is bounded we note that $(x A)'<1-x^2 A {f'}^2$.
Choose $\epsilon>0$ such that $A(x)>A(\bar x)/2$ for
$\bar x-\epsilon<x<\bar x$; then
$(x A)'<1-(\bar x-\epsilon)^2 A(\bar x){f'}^2/2$
and integrating from $\bar x-\epsilon<x$ gives that
$\int_{\bar x-\epsilon}^x f'(x)^2 < \infty$ and hence, by the
Cauchy-Schwarz inequality,
$\int_{\bar x-\epsilon}^x |f'(x)| <\infty$.
Thus, $f$ is bounded. This implies by Eq.(2.14c) that
$(\ln C)'$ is also bounded so both $C$ and $1/C$ are bounded. Now, (2.14a)
says that $x^2 f'/C$ is bounded so $f'$ is bounded.
\end{proof}
\begin{rem} It follows from Lemma~2.3 that the only
possible
obstruction to extendability
of $a$-orbits to arbitrarily large $x$ is
$\lim_{x \rightarrow \bar x}A=0$ for some $\bar x$.
If that happens we will say that the
solution {\em crashes} at $\bar x $.
\end{rem}
Let us define the
function $g=1-AC^2$. The following two properties of this function
will play an important role in our
discussion.
\begin{lem} We have
\begin{description}
\item[(a)] If $g(x)\leq 0$, then $g'(x)>0$;
\item[(b)] If $g(x_0)\geq 0$, then $g(x)>0$ for all $x>x_0$.
\end{description}
\end{lem}
\begin{proof} A simple calculation yields
\begin{equation}
g'=C^2 \left(\frac{1-A}{x} + x A {f'}^2 - x g f^2\right).
\end{equation}
The part (a) follows immediately from (2.18). To prove the part (b)
note that $g'(x_1)>0$ if $g(x_1)=0$, so $g$ cannot cross zero from
above. \end{proof} \vskip 0.15cm The restriction $\alpha>1$ in
Theorem~2.1 can be easily seen as follows. Suppose that there is a
globally regular solution with $\alpha \leq 1$. Since
$g(0)=1-\alpha^2$, it follows from Lemma~2.4 that $g(x)$ is positive
for all $x$.  Multiplying Eq.(2.14a) by $f$ and integrating by parts
we get that $f f'>0$ for all $x$, hence $f^2$ is monotone increasing
which is obviously impossible for globally regular solutions (in fact
such solutions crash at finite $x$ as follows easily from
Eq.(2.14b)). Thus we have
\begin{lem}
There are no (nontrivial) globally regular solutions for
$\alpha \leq 1$.
\end{lem}
Note that Lemma~2.5 implies in particular that there are no static
($\alpha=0$) globally regular solutions. In view of Lemma~2.5 from
now on we always assume that $\alpha>1$.
\begin{defn} The rotation function $\theta(x,a)$ of an $a$-orbit is defined
by $\theta(0,a)=0$, $\tan{\theta(x,a)}=-f'(x)/f(x)$ and $\theta(x,a)$
is continuous in $x$.
We will drop the second argument of $\theta$ if there is no danger
of confusion.
\end{defn}
Now we list the basic properties of the rotation function of
$a$-orbits which we will need below.
\begin{lem} For any nonnegative integer $n$ we have:
\begin{description}
\item[(a)] If $\theta(x_1) > (n+1/2)\pi$ for some $x_1$, then
$\theta(x) > (n+1/2)\pi$ for all $x>x_1$.
\item[(b)]  If $\theta(x_1) < n\pi$ for some $x_1$ and $g(x_1)\geq 0$,
then $\theta(x) < n\pi$ for all $x>x_1$.
\item[(c)] There are at most two values of $x$ with
$\theta(x)=n\pi$.
\end{description}
\end{lem}
\begin{proof} (a) We note that $\theta'(x)=({f'}^2-f f'')/(f^2+{f'}^2)$, so
$\theta'(x)=1$ if $\theta(x)=(n+1/2)\pi$.\\ (b) If $x>x_1$ and
$g(x_1) \geq 0$, then $g(x)>0$ by Lemma~2.4. Next, we note that
$\theta'(x)=-g(x)/A(x)<0$ if $\theta(x)=n\pi$ and $g(x)>0$. If
$g(x)=0$ then $\theta'(x)=0$ but $\theta''(x)=-g'(x)/A(x)<0$ since
$g'(x)>0$ when $g(x)=0$ by Eq.(2.18).\\ (c) The function
$\theta(x)-n\pi$ changes sign at each zero for which $g(x) \ne 0$.
From Lemma~2.4, $g$ changes sign at most once. Thus, for $n>0$,
$\theta(0)-n\pi<0$ and at $x_1$, the first zero of $\theta(x)-n\pi$,
if $g(x_1)\geq0$ then by part b) $\theta(x)-n\pi<0$ for all $x>x_1$.
If $g(x_1)<0$ then $\theta(x)-n\pi$ changes sign at $x_1$, and hence,
at $x_2$, the next zero of $\theta(x)-n\pi$, $g(x_2) \geq 0$ and
hence $\theta(x)-n\pi<0$ for all $x>x_2$. For $n=0$,
$\theta(0)-n\pi=0$, $\theta(x)>0$ near $x=0$ and if $\theta(x_1)=0$
then $g(x_1) \geq 0$, hence, $\theta(x)<0$ for all $x>x_1$.
\end{proof}
Before going into details of Sections~3--8, let us outline the main
idea of the proof of Theorem~2.1. According to this theorem there
exists a countable family of globally regular solutions distinguished
by nodal class. We first show (section 3) that there is a continuous
one-parameter family of local solutions depending on $a = f(0)$; we
all these solutions $a$-orbits. In Section~6 we show that an
$a$-orbit that has bounded rotation and that is defined for all $x$
is a globally regular solution, that is, it has the correct
asymptotic behavior as $x \rightarrow \infty$. The existence of
$a$-orbits with bounded rotation that are defined for all $x$ is
proven in each nodal class  by an inductive application of a shooting
argument. The zeroeth solution we construct has
$\theta(x,a_0) < \pi/2$ for all $x$; the first solution has
$\theta(x,a_1) < 3 \pi/2$
(and greater than $\pi/2$ for large $x$), etc. This is shown in
Fig.~1. The crucial step of our argument is the control of behavior
of $a$-orbits for large and small values of the parameter $a$. In
Section~4 we show that for sufficiently small $a$ the $a$-orbit has
arbitrarily large rotation; more precisely, there is a number $b_n$
such that $\theta(x,a) > n \pi$ for some $x$ if $a < b_n$.  In
contrast, we show in Section~5 that for
$a >> 1$ the $a$-orbit exits Q4 directly to Q1 (see Fig.~1).

Now, to prove the existence of a globally regular solution in the
zeroeth nodal class we let
$a_0= \inf \{a|\: \theta(x,a) < \pi/2$
for all $x$ for which the $a$-orbit is defined\}.
Note that $a_0 \geq b_1 > 0$. We then prove that the $a_0$-orbit is
the globally regular solutions in the zeroeth nodal class. It is
clear that the $a_0$-orbit has rotation $\theta(x,a_0)\leq \pi/2$ for
otherwise all nearby orbits would have rotation $> \pi/2$ which
contradicts the definition of $a_0$.
It is also easy to see that the $a_0$-orbit cannot exit Q4 to Q1 because
again, nearby orbits would also do so which contradicts the
definition of $a_0$. Hence, the $a_0$-orbit must stay in Q4; it
either crashes or is defined for all $x$ and is a globally regular
solutions in the zeroeth nodal class. Thus, it remains to show that
the $a_0$-orbit does not crash. The (technical) crash lemma of
Section~7 shows that if an orbit crashes in Q4 then nearby orbits
either crash in Q4 or exit Q4 to Q1. Thus the $a_0$-orbit cannot
crash because nearby orbits would all be in
$\{a|\theta(x,a)<\pi/2$ for all $x$ for which the $a$-orbit is defined\}
 and $a_0$ would not be the infimum of that set.

To show the existence of globally regular solutions in higher nodal
classes we proceed as above. We let
$a_n= \inf \{a|\theta(x,a) <(n+1/2)\pi$ for all $x$ for which the $a$-orbit is defined\}.
We then show that $\theta(x,a_n)<(n+1/2)\pi$.
We again use the crash lemma as we did in the $n = 0$ case to show
the $a_n$-orbit is defined for all $x$. The only difference is that
we must show that $\theta(x,a_n)> n \pi$. That fact follows
easily from lemmas 2.6b and 6.3. \vskip -0.8cm
\begin{figure}[!ht]
\centering
\includegraphics[width=1.25\textwidth,bb= 85 75 446 327]{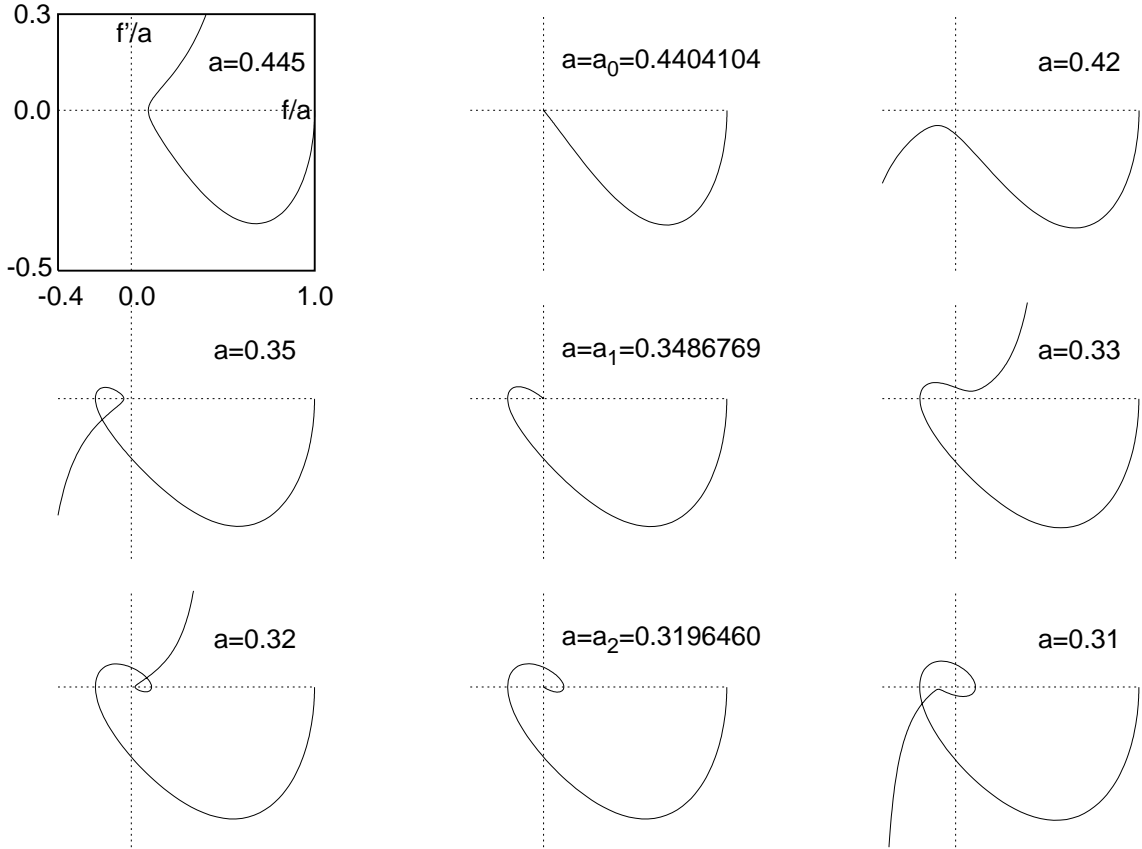}
\caption{The projection of $a$-orbits on the $(f,f')$ phase plane for
several selected  values of the shooting parameter $a$.}
\end{figure}
\section{Local existence}
\begin{prop}
There exists a two-parameter family of local solutions of Eqs.(2.14)
near $x=0$ satisfying the initial conditions (2.17).
\end{prop}
\begin{proof} The proof is standard so we just outline it.
We introduce new variables $w=f'$, $z=\ln(C)$, $B=(1-A)/x$, and
rewrite Eqs.(2.14) as the first order system
\begin{subequations}
\begin{eqnarray}
f'&=&w,\\
(x^2 w)'&=&\frac{x^2}{A}\left((x f^2 -B)w+ f(1-AC^2)\right),\\
(x^2 B)'&=& x^2 (A w^2 + f^2 A C^2 + f^2),\\
z'&=&\frac{x f^2 - B}{A}.
\end{eqnarray}
\end{subequations}
We will use the sup norm throughout this discussion: $\norm{h}$
means the $\sup\{|h(x)|: 0\leq x \leq r\}$.

Consider the space $X$ of quadruples of functions $(f,y,B,z)$ where
$\norm{f-a}\leq 1, \norm{w}\leq 1, \norm{B}\leq M$, and
$\norm{z-\ln(\alpha)}\leq 1$ and each of the four functions is in
$C^0([0,r])$, the space of continuous functions defined on the
interval $0\leq x\leq r$ with the sup norm. $X$ is a complete metric
space if we take as metric the maximum of the four components. We
define a map $T: X \rightarrow X$ by $T(f,w,B,z)=(T_1,T_2,T_3,T_4)$
where
\begin{subequations}
\begin{eqnarray}
T_1 &=& a + \int_0^x w \;ds,\\
T_2 &=& \frac{1}{x^2} \int_0^x \frac{s^2}{A} \left((s f^2 -B) w + f
(1-A C^2)\right) ds,\\
T_3 &=& \frac{1}{x^2} \int_0^x s^2 \left( A w^2 + f^2 A C^2 +f^2)\right) ds,\\
T_4 &=& \ln{\alpha} + \int_0^x \frac{1}{A} ( s f^2 -B) ds.
\end{eqnarray}
\end{subequations}
One verifies easily that $T$ does in fact take $X$ to $X$ and that
$T$ is a contracting map if $r$ is sufficiently small, and that a
fixed point of $T$ is a solution to our equations. The proof that the
solution depends continuously on $a$ is also routine.
\end{proof}

\section{Behavior of solutions for small $a$}
In this section we show that the rotation $\theta(x,a)$ of the
$a$-orbit is
arbitrarily large if $a$ is sufficiently small
and $x$ is sufficiently large.
\begin{prop}
For any $n>0$, there exists a $b_n$ such that for $a<b_n$ there is an
$x$ with $\theta(x,a)>n \pi$.
\end{prop}
\begin{proof} Let $\tilde f=f/a$. Then, Eqs.(2.14) become
\begin{subequations}
\begin{eqnarray}
\left(\frac{x^2 \tilde f'}{C}\right)' &=& \frac{x^2}{AC} (1-AC^2)
\tilde f,\\
A' &=& \frac{1-A}{x}-a^2 x ( A \tilde {f'}^2+A C^2 \tilde f^2
+ \tilde f^2),\\
{C'} &=& \frac{C}{x A}(A -1+ a^2 x^2 \tilde f^2)
\end{eqnarray}
\end{subequations}
with the behavior at the origin
\begin{equation}
\tilde f(0)=1, \quad A(0)=1, \quad C(0)=\alpha.
\end{equation}
For $a=0$ (decoupling of gravity) Eqs.(4.1bc) with conditions (4.2)
have constant flat-spactime solutions $A \equiv 1$, $C \equiv
\alpha$. Inserting these solutions into Eq.(4.1a) gives the Bessel
equation
\begin{equation}
(x^2 \tilde f')' + x^2 (\alpha^2-1) \tilde f =0,
\end{equation}
whose unique solution satisfying (4.2) is
\begin{equation}
\tilde f(x) =
\frac{\sin{\sqrt{\alpha^2-1}\,x}}{\sqrt{\alpha^2-1}\,x}.
\end{equation}
This solution has infinite rotation as
$x \rightarrow \infty$. If
$x > \frac{\ {n\pi }}{\sqrt{\alpha ^2-1}}$
 then $\theta(x,0) > n \pi $ so for $a$ close to $0$, say
$a < {b_n}$, we have $\theta(x,a) > n\pi$ because solutions of
Eqs.(4.1) are continuous in $a$ and $x$. This concludes the proof of
Proposition~4.1.
\end{proof}
\section{Behavior of solutions for large $a$}
\begin{prop}
The $a$-orbits with sufficiently large $a$ exit Q4 directly to Q1.
\end{prop}

We define new variables
\begin{equation}
y=ax, \quad \tilde v(y)=a (f(x)-a), \quad \tilde A(y)=A(x), \quad
\tilde C(y)=C(x).
\end{equation}
Then, Eqs.(2.14) become (where now the prime denotes the derivative
with respect to $y$)
\begin{subequations}
\begin{eqnarray}
\left(\frac{y^2 v'}{\tilde C}\right)' & = & \frac{y^2}{\tilde A
\tilde C} (1-\tilde A \tilde C^2) (1+ \frac{\tilde v}{a^2}),\\
\tilde A' &=& \frac{1-\tilde A}{y} - y \left(\frac{1}{a^2} A
{\tilde {v'}}^2 + (1+\tilde A {\tilde C}^2)\left(1+\frac{\tilde
v}{a^2}\right)^2\right),
\\
\tilde C' &=& \frac{\tilde C}{y \tilde A} \left(\tilde A -1 + y^2
(1+\frac{\tilde v}{a^2})^2\right).
\end{eqnarray}
\end{subequations}
The initial conditions at $y= 0$ are
\begin{equation}
\tilde A(0)=1, \quad \tilde C(0)=\alpha>1, \quad \tilde v(0)=0, \quad
\tilde v'(0)=0.
\end{equation}
As $a \rightarrow \infty$, the solutions of Eqs.(5.2) tend uniformly
on compact intervals to the solutions of the following limiting
system
\begin{subequations}
\begin{eqnarray}
\left(\frac{y^2 v'}{C}\right)' & = & \frac{y^2}{A C} (1- A C^2),\\
A' &=& \frac{1-A}{y} - y (1+A C^2),\\
C' &=& \frac{C}{y  A} (A -1 + y^2).
\end{eqnarray}
\end{subequations}
satisfying the same initial conditions as in (5.3). The rest of this
section is devoted to the analysis of Eqs.(5.4). Our goal is to show
that $v'(y)$ becomes positive at a point $y_1 < \bar y$. This would
imply that $v(y)$ is bounded below, i.e., there is $d>0$ such that
$v(y)>-d$ for $y<y_1$, and therefore $\tilde v(y)>-d-1$ if $a$ is
sufficiently large. Then $f(x)>a-(d+1)/a$ and $f'(x)>0$ for
$x=y_1/a$, hence, if $a>\sqrt{d+1}$, the $a$-orbit exits Q4 to Q1
directly without entering Q3.

Note that the function $v$ decouples from Eqs.(5.4bc) for
the metric coefficients -- this fact  considerably simplifies the
analysis.
\begin{lem}
The solution of Eqs.(5.4) crashes at some $\bar y$, that is,
$A(\bar y)=\lim_{y\rightarrow\infty} A(y)=0$. Moreover, $1<\bar y<\sqrt{3}$.
\end{lem}
\begin{proof}

Note that $(y A)'<1-y^2$ so integrating gives
$A<1-y^2/3$. Therefore, $A(\bar y)=0$ for $\bar y <\sqrt{3}$. To show
that $\bar y>1$ assume $\bar y \leq 1$.  Then
$(\frac{y}{C})'=\frac{1-y^2}{AC}\geq 0$ so if $0<\tau<y<\bar y$ we
have $y/C(y)>\tau/C(\tau)$ or $C(y)<C(\tau)/\tau$ so $C$ is bounded.
Since $(AC)'=-y A C^3$, $(\ln(AC))'=-y C^2$ is bounded below. Thus,
by integrating one concludes that $\lim_{y \rightarrow \bar y} AC(y)>0$.
But $(AC)^2 =A (A C^2); AC^2\leq \alpha^2$ and
$\lim_{y\rightarrow \bar y} A(y)=0$ so
$\lim_{y \rightarrow \bar y} (AC(y))^2=0$.
This contradicts $\lim_{y \rightarrow \bar y} AC(y)>0$
so we must have $\bar y >1$.
\end{proof}

\begin{proof}[Proof of Proposition~5.1] In order to prove that $v'(y)$
becomes positive at some point $y_1 < \bar y$, we will show that
$v'(\bar y)>0$. By Eq.(5.4a) we have $v'(y)= \frac{C}{y^2} \int_0^y
\frac{z^2 (1-A C^2)}{A C} dz$, so we must show that $\int_0^{\bar y}
\frac{y^2 (1-A C^2)}{A C} dy >0$.

The proof of this fact is
divided into two cases: (i) $\bar y^2 \geq 3/2$,
and (ii) $\bar y^2 <3/2$.
Before considering these cases we list some useful properties
of the function $g=1-A C^2$.
\begin{lem} We have:
\begin{description}
\item[(a)] $g'= (1- A -y^2 g) C^2/y$;
\item[(b)] if $g(y_0)\geq 0$, then $g(y)>0$ for all $y>y_0$;
\item[(c)] $g'>0$ if $g \leq 1/3$;
\end{description}
\end{lem}
\begin{proof} Part (a) is a calculation. For (b) note that $g'(g=0)>0$ so $g$
cannot cross zero from above. For (c) we have $(y A)'<1-y^2$ so
integrating gives $1-A>y^2/3$ and hence, $g'>y(1/3-g) C^2$.
\end{proof}
We return now to the proof that
$\int_0^{\bar y} \frac{y^2 g}{A C} dy>0$.
We first consider the case (i)~$\bar y^2 > 3/2$.

A calculation shows that $y^2 g= (2 y^3/3+y A -y)'$, hence
$\int_0^{\bar y} y^2 g dy =2 \bar y^3/3 - \bar y >0$
if $\bar y^2 > 3/2$. Since $g(0)=1-\alpha^2<0$, this implies that
$g(\sigma)=0$ for some $\sigma <\bar y$ and therefore  $g(y)>0$ for
$y>\sigma$.  Note that $AC$ is monotone decreasing because
$(AC)'=-yA C^3 <0$. Thus
\begin{equation}
\frac{y^2 g(y)}{A(y)C(y)} \geq \frac{y^2 g(y)}{A(\sigma)C(\sigma)}
\quad \text{for} \quad  0\leq y\leq \bar y
\end{equation}
and
therefore
\begin{equation}
\int_0^{\bar y} \frac{y^2 g}{A C} dy \geq \frac{1}{A(\sigma)
C(\sigma)} \int_0^{\bar y} y^2 g dy > 0.
\end{equation}
\end{proof}
Now we consider the case (ii) $\bar y^2 \leq 3/2$.
\begin{lem}
Define the function $p=1+y^2 g -y^2$. If
$y^2\leq 3/2$, then $p(y)>0$.
\end{lem}
\begin{proof} Note that $p(0)=1$. Let $y_1$ be the first zero of $p$,
that is, $p(y_1)=0$ and $p'(y_1)\leq 0$. If $g(y_1)> 1/3$ then
$p=y^2 g + 1-y^2 > y^2/3 + 1-y^2=1-2y^2/3$. Thus $p$ can have a zero for
$y_1^2 \leq 3/2$ only if $g(y_1)\leq 1/3$. Then, from Lemma~(5.3),
$g'(y)>0$ for all $y\leq y_1$. Define a function $k(y)=2-3 A -y^2$. A
calculation gives $y^3 g'=(y(k+p))'$ so by integrating we get
$k(y_1)>0$. On the other hand we have $p'=y C^2 (k-p)$, so $k(y_1)\leq 0$;
contradiction.
\end{proof}
To show that $\int_0^{\bar y} \frac{y^2 g}{A C} dy >0$, we rewrite it as
\begin{equation}
\int_0^{\bar y} \frac{y^2 g}{A C} dy = \int_0^{\bar y}
\frac{p-1+y^2}{A C} dy  =\int_0^{\bar y} \frac{p}{A C} dy -
\int_0^{\bar y} \frac{1-y^2 }{A C} dy.
\end{equation}
The first term on the right hand side of (5.7) is positive because
$p$ is positive. To compute the second term, note that
\begin{equation}
(\frac{y}{C})'=\frac{1-y^2}{AC},
\end{equation}
 hence $L=\lim_{y\rightarrow \bar y} (y/C)$ exists and is finite since
$\bar y > 1$ by Lemma~5.2. If $L > 0$ then $\lim_{y\rightarrow
\bar y} C(y)=\bar y/L<\infty$ so $C$ is bounded. Since
$\lim_{y\rightarrow \bar y} A(y)=0$ we conclude that
$\lim_{y\rightarrow \bar y} AC(y)=0$. But $(\ln AC)'=y C^2$ is
bounded so $\ln AC$ is bounded below and hence $\lim AC \neq 0$.
This contradiction shows that $L=0$. Thus, the second term on the
right hand side of (5.7) is zero. This concludes the proof of
Proposition~5.1.
\section{Asymptotics of globally regular solutions}
In this section we derive the leading asymptotic behavior of
globally regular solutions.  We use $\lim$ to denote $\lim_{x
\rightarrow \infty}$.
\begin{prop}
An $a$-orbit which exists for all $x$ and has bounded rotation is
asymptotically flat. The leading asymptotic behavior for
$x\rightarrow \infty$ is
\begin{equation}
A(x) \sim 1-\frac{2 M_{\infty}}{x}, \qquad
C(x) \sim C_{\infty} e^{\frac{2M_{\infty}}{x}},
\qquad f(x) \sim f_{\infty} e^{-b x},
\end{equation}
where $0<M_{\infty}<\infty$, $0<C_{\infty}<1$, and
$b=\sqrt{1-C^2_{\infty}}$.
\end{prop}
To prove this proposition we need several partial results.
\begin{lem}
An $a$-orbit which exists for all $x$ and has bounded rotation is
ultimately in the second (Q2) or fourth (Q4) quadrant.
\end{lem}
\begin{proof}  If $\theta(x)$ is bounded above then there is an
 integer $n \geq 0$ such that $\theta(x)< (n+1/2)\pi$ for all $x$ but
$\theta(x_1)>(n-1/2)\pi$ for some $x_1$ and hence, by Lemma~2.6a for
all $x>x_1$, $(n-1/2)\pi<\theta(x)<(n+1/2)\pi$. We next show that
there is an $x_2$ such that for all $x>x_2$, $n\pi<\theta(x)<(n+1/2)\pi$
(that is, the orbit is ultimately in Q2 or Q4). Note that, by
Lemma~2.6c the orbit must satisfy either $n\pi<\theta(x)<(n+1/2)\pi$
or $(n-1/2)\pi<\theta(x)<n\pi$, that is the orbit
 must lie in Q3 or Q2 if $n$ is odd and in Q1 or Q4 if $n$ is even.
We must rule out the possibility that the orbit is in Q1 or Q3.
Assume
that the orbit lies in Q1 or Q3 for all $x>x_1$. Then $f(x) f'(x)>0$
for all $x>x_1$, so $f^2(x)\geq f^2(x_1)$ for all $x>x_1$. From Eq.(2.15b) we
have $ (xA)'=1-x^2 A{f'}^2-x^2 f^2 A C^2 - x^2 f^2$ so
$(xA)'<1-x^2 f^2<1-x^2 f^2(x_1)$ and hence $A$ goes to zero in finite $x$.
This contradiction concludes the proof.
\end{proof}
\begin{lem}
Under the assumptions of Proposition~6.1 the function $g=1-A C^2$ is
eventually positive.
\end{lem}
\begin{proof}
Suppose that $g(x)\leq 0$ for all $x$. We claim that this implies
$\lim A=1$. To see this, suppose that $\liminf A=1-4 \epsilon$ for
some $\epsilon>0$. Let $-\beta=\lim g \leq 0$ which exists because
$g'>0$. Note that $g(x)<-\beta$ for all $x$. Choose an $x_1$ such
that $g(x_1)>-\beta-\epsilon$. If $A(x_2)<1-3\epsilon$ for some
$x_2>x_1$, then by (2.18)
$g'(x)>C^2(1-A)/x>(1-A)/x=x(1-A)/x^2>x_2(1-A(x_2))/x^2>3\epsilon
x_2/x^2$ for $x>x_2$, where the last but one inequality follows from
the fact that $x(1-A(x))$ is monotone increasing. Integrating this
inequality from $x_2$ to $2 x_2$ say, we get $g(2 x_2) > g(x_2)+
3\epsilon/2 > -\beta-\epsilon+3 \epsilon/2>-\beta$; contradiction.
Thus, $\liminf A=1$ and hence $\lim A=1$. Since $\lim g=\lim(1-AC^2)$
exists, $\lim C$ also exists and is finite. Next, from Lemma~6.2 we
know that the $a$-orbit is ultimately in Q2 or in Q4. For
concreteness we consider the case of Q4 (the proof of the Q2-case is
identical), that is $f(x)>0$ and $f'(x)<0$ for sufficiently large
$x$. Then, from (2.14a), $\lim (x^2 f'/C)$ exists so $\lim (x^2
f')=-\tau<0$ exists as well (where $\tau$ might be infinite; the
point is that $\tau \neq 0$). Now, by L'H\^{o}pital's rule, $\lim x
f= -\lim (x^2 f') =\tau$. But (2.14c) says $(\ln C)'>\tau^2/4x$ which
implies $\lim C=\infty$, a contradiction.
\end{proof}
\begin{proof}[Proof of Proposition~6.2]

From the previous lemma we know that there exists an $x_1$ such
that $g(x)>0$ for $x>x_1$. Let $u=A C f/g$ for $x>x_1$. A calculation
shows that $u'= -A C (f C^2 (1-A)/x - f' g + x f {f'}^2)/g^2$ so
$u'<0$ if $g>0$. Multiplying Eq.(2.14a) by $u$ we obtain
\begin{equation}
(x^2 A f f'/g)' = x^2 f^2 + x^2 f' u'/C.
\end{equation}
The right hand side is positive for $x>x_1$ so $x^2 A f f'/g$ is
negative and increasing, hence it has a finite non-positive limit.
This implies that $x^2 f^2$ is integrable. Similarly, multiplying
Eq.(2.14a) by $f$ we obtain
\begin{equation}
x^2 f f'/C =(x^2 f^2 g + A x^2 {f'}^2)/(AC).
\end{equation}
The right hand side is positive for $x>x_1$ so $x^2 f f'/C$ is
negative and increasing, hence it has a finite non-positive limit.
This implies that
$A x^2 {f'}^2$ is integrable (recall that $A C$ is
monotone decreasing). The integrability of $x^2 f^2$ and $A x^2
{f'}^2$ implies via Eq.(2.15) that $\lim M=M_{\infty}<\infty$ exists.
This concludes the proof that $A(x) \sim 1-2 M_{\infty}/x$.

Having $\lim A=1$ we can strengthen Lemma~6.3 by showing that $\lim
g=g_{\infty}>0$ exists. To see this choose an $x_1$ such that
$g(x_1)>0$. Then $A C^2(x_1)<1$, hence $A C(x_1)<1$. Since $A C$ is
monotone decreasing, we have  $A C(x)< A C(x_1)$ for $x>x_1$ and thus
$\lim A C<1$. Hence, $\lim A C^2 = (\lim A C)^2/\lim A<1$. Since
$g=1-A C^2$, $\lim g$ exists and $\lim g>0$.

Now we have all we need to derive the asymptotics of $f$. Let
$r=f'/f$. Then $r'=f''/f-r^2 = -r(1+A-x^2 f^2)/(x A) +g/A =
g_{\infty}-r^2 +\epsilon(x)$, where $\lim \epsilon=0$. Let
$\sigma(x_2)=\max (|\epsilon(x)|)$ for $x>x_2$ and assume that $x_2$
is sufficiently large so that $g_{\infty}>\sigma(x_2)$. If
$r(x_2)>-\sqrt{g_{\infty}-\sigma(x_2)}$, then clearly $r$ becomes
eventually positive which contradicts that the orbit is eventually in
Q2 or Q4. If $r(x_2)<-\sqrt{g_{\infty}+\sigma(x_2)}$, then $\lim
r=-\infty$ -- this is impossible because then by L'H\^{o}pital's rule
$\lim r=\lim f''/f'=\lim g/r=0$. Therefore $r(x_2)$ must be
sandwiched in the interval
$-\sqrt{g_{\infty}+\sigma(x_2)}<r(x_2)<-\sqrt{g_{\infty}-\sigma(x_2)}$.
Since $x_2$ is arbitrarily large and $\lim \sigma=0$, we conclude
that $\lim r=-\sqrt{g_{\infty}}$. The asymptotics of $f$ given in
(6.1) follows immediately from this.

Finally, inserting the derived leading asymptotic behavior of $A$ and
$f$ into Eq.(2.14c), we obtain $C'/C \sim -2 M_{\infty} /x$, from
which the asymptotics of $C$ follows trivially.
\end{proof}
\section{Solutions that crash}
\begin{prop}
If the $b$-orbit crashes at some $\bar x$ then $g(x)>0$ for $x$ near
$\bar x$.
\end{prop}
\begin{proof}

Suppose that
$g(x)<0$ for all $x< \bar x$, so $A C^2(x)>1$ for all $x<\bar x$. We
have  from (2.18) that $g'>A C^2 x {f'}^2>x {f'}^2$. Integrating this
inequality from some $x_1>0$ to some $x_2<\bar x$, we obtain
\begin{equation}
x_1 \int_{x_1}^{x_2} {f'}^2 dx < \int_{x_1}^{x_2} x {f'}^2 dx <
g(x_2)-g(x_1)<\alpha^2-1,
\end{equation}
which implies (by the Cauchy-Schwartz inequality) that $f$ is
bounded.

Next, $A(\bar x)=0$, $A C^2>1$, implies that $\lim_{x \rightarrow
\bar x^-}C =\infty$; moreover,  by (2.14c) $(\ln C)'< x f^2/A$, hence
$x f^2/A$ is not integrable near $\bar x$. Since $f$ is bounded, this
shows that $1/A$ is not integrable near $\bar x$. But from (2.18),
$g'>C^2 (1-A)/x= A C^2 (1-A)/(x A)>1/(2 x A)$, so $g'$ is not
integrable near $\bar x$, which contradicts the fact that $g$ is a
bounded function.
\end{proof}
The importance of Proposition~7.1 derives from Lemma~2.6b which says
that if $g>0$ then rotation stops. The main result of this section is
the crash theorem which states that if an orbit has bounded rotation
and crashes, then nearby orbits also have similarly bounded rotation.
The precise statement is given in Proposition~7.2. Since we consider
more than one orbit in this section, we use the notation $A(x,a)$ to
denote the value of $A$ at $x$ for the $a$-orbit, etc.
\begin{prop}[Crash Theorem]  If the $b$-orbit crashes at $x=\bar x$
and
\begin{description}
\item[(a)]
if $(k-1/2)\pi<\theta(x,b)<k\pi$, $k\geq 1$, for $x$ near $\bar x$,
then nearby orbits have rotation $<k\pi$ for $x\geq \bar x$;
\item[(b)] if
$k\pi<\theta(x,b)<(k+1/2)\pi$, then nearby orbits have rotation
$<(k+1/2)\pi$.
\end{description}
\end{prop}
\begin{proof} Part (a): Suppose the $b$-orbit crashes in Q3 or Q1. By
Proposition~7.1, $g(x_1,b)>0$ for some $x_1<\bar x$ with
$(k-1/2)\pi<\theta(x_1,b)<k\pi$; hence, for $a$ sufficiently near $b$
we have $g(x_1,a)>0$ with $(k-1/2)\pi<\theta(x,a)<k\pi$. By
Lemma~2.6b, $\theta(x,a)<k\pi$ for all $x>x_1$.
Part (b): This case is much more difficult and will require several
auxiliary results. It follows from part (a) that nearby orbits have
rotation $<(k+1)\pi$; we must prove a much more difficult result,
namely that nearby orbits have rotation $<(k+1/2)\pi$.
\end{proof}
\begin{rem}
It is clear from numerical observations that no $a$-orbit crashes in
Q2 or Q4; however, that appears to be quite difficult to prove.
Moreover, one can easily construct orbit segments that start, for
example, at $x=1$ with $f=5, f'=0, A=0.2, C=3$, say, that crash in
Q4. Such orbit segments have
$\lim_{x \rightarrow \bar x^-}f'(x)=-\infty$.
Nevertheless, the next lemma shows that $A {f'}^2$
remains bounded at crash.
\end{rem}
\begin{lem}
If an $a$-orbit is defined for $x<x_2, f f'(x)<0$ for $x_1<x<x_2$,
$f^2(x_1)<B$, and $f'(x_1)=0$ then $A {f'}^2(x)\leq\max(B,\alpha^2/3)$.
In particular, if an orbit crashes in Q2 or Q4,
$\lim_{x \rightarrow \bar x^-} A(x)f'(x)=0$.
\end{lem}
\begin{proof}
We set $q=A {f'}^2$ and then compute
that
\begin{equation}
x q'= -(3+x^2 {f'}^2 + x^2 C^2 f^2) q -{f'}^2 + 2 x f f' +x^2 f^2
{f'}^2 - 2 A C^2 x f f'.
\end{equation}
Note that $q\geq 0$ and all terms on the right side of (7.3) are
negative except for the last two. If $q>B$, we combine the term
$-q x^2 {f'}^2$ with $x^2 f^2 {f'}^2$; clearly,
$x^2 f^2 {f'}^2-q x^2{f'}^2=(f^2-q) x^2 {f'}^2 \leq 0$.
Next, we combine the term $-q x^2{f'}^2 C^2$ with
$-2 x f f' A C^2$ to get $-A C^2 (y^2-2y)$ where
$y=-x f f'$; the maximum value of this expression  occurs when $y=1$
and that value is $A C^2\leq \alpha^2$ by Lemma~2.4. Hence, if
$q\geq\alpha^2/3$, then
$-q (x^2 {f'}^2 C^2) - 3 q -2 x f {f'}^2 A C^2 \leq 0$.
Thus, $q\geq \max(B,\alpha^2/3)$ implies that $q'<0$;
consequently, $A {f'}^2(x)\leq \max (B,\alpha^2/3)$.
Since $A A{f'}^2=(A {f'})^2$, and $A {f'}^2$ is bounded and
$\lim{x\rightarrow\bar x^-} A(x)=0$,
$\lim_{x\rightarrow \bar x^-} (A(x) f'(x))^2=0$,
hence $\lim_{x\rightarrow \bar x^-} A(x) f'(x)=0$.
\end{proof}
We can now discuss the strategy of the proof of part (b) of
Proposition~7.2. We want to show that if an orbit is sufficiently
close to an orbit that crashes in Q4 then it must either crash or
exit Q4 to Q1 (the case in which the orbit crashes in Q2 is
completely symmetric). To that end, let $v(x)=A(x) f'(x)$. We will
prove that $v(x,a)$ goes to $0$ if $a$ is sufficiently close to $b$
and $f(x,a)>0$. This means either $f'=0$ and hence the orbit is
exiting Q4 to Q1, or $A=0$, that is, the orbit is crashing in Q4.
Note that $v'(x)=-(2 A f'- x f + x A C^2 f + x^2 A {f'}^2 + x^2 f^2 f' A C^2)/x
=-v (2+x^2{f'}^2 + x^2 f^2 C^2)/x+f g > f g$. We know
that $v(x,b)$ goes to $0$ at crash so nearby orbits will also have
$v$ small for $x$ near $\bar x$. We will show that $f$ and $g$ are
both uniformly bounded away from $0$ in an interval about $\bar x$.
That is, the size of the interval and the bounds work for all $a$
near $b$. That is enough to force $v$ positive. The most technical
part of the proof involves showing that nearby orbits stay in Q4 long
enough to have $v$ go positive. Since $f'$ goes to $-\infty$ at
crash, nearby orbits have $f'$ large also. Now, (2.14a) can be
written as $x A f'' + (1+A-x^2 f^2) f' - x g f=0$; moreover, to get
to Q3  orbits must pass through $x f(x)<1$ which means that the
coefficient of $f'$, $(1+A -x^2 f^2)$, is positive. That is enough to
bound $f'$.

The details of the proof, especially Lemma~7.5, are tedious. We will
restrict ourselves to an interval $0.99\: \bar x < x < 1.01\: \bar x$
and replace $x$ by $\bar x$ (whenever justified) in making estimates.

We show next that if the $b$-orbit crashes at $x=\bar x$ with
rotation $k \pi<\theta(x,b) < (k+1/2)\pi$ then
$|\bar x f(\bar x)|\geq 1$.
\begin{lem}
If the $b$-orbit crashes at $x=\bar x$ with $\theta(x,b)<(k+1/2)\pi$
for all $x <\bar x$ and $\theta(x,b)>k \pi$ for $x$ near $\bar x$,
then  $|\bar x f(\bar x)| \geq 1$, in particular $f(\bar x)\neq 0$.
\end{lem}
\begin{proof}
The assumption on $\theta(x,b)$ tells us that the orbit lies in Q2 or
Q4 for $x$ near $\bar x$. For simplicity of exposition we only
discuss the case of Q4, i.e., $f(x) \geq 0, f'(x) \leq 0$.
In particular, $f$ is a monotone function and hence has a limit at
$\bar x$. Thus, $h(x)=x f(x)$ is continuous; in particular, if we suppose
that $\bar x f(\bar x)<1$, then $h(x)<1$ for $x$ near $\bar x$. Since
$A(\bar x)=0$, we get from (2.14c) that $x A C'=C (A-1+x^2 f^2)<0$
for $x$ near $\bar x$. We conclude that $C$ is bounded above, hence
$\lim_{x \rightarrow \bar x^-} A C^2=0$ and
$\lim_{x \rightarrow \bar x^-} g=1$. Since $g>0$,
the right hand side of Eq.(2.14a) is positive
and hence $x^2 f'/C$ is bounded and since $C$ is bounded we conclude
that $f'$ is bounded; thus
$\lim_{x \rightarrow \bar x^-} A{f'}^2=0$. Then, from (2.14b),
$x A'=1-A -x^2 f^2 - x^2 (A {f'}^2 + AC^2 f^2)$, we see that
$A'>0$ near $\bar x$ so there is no crash.
This is a contradiction so we conclude that $\bar x f(\bar x) \geq 1$
and hence $f(\bar x)>0$.
\end{proof}
\begin{lem}
There is a $\gamma>0$ such that $h(x,a)=x f(x,a)>1/4$ for all $a$
sufficiently near $b$ and $\bar x <x<\bar x +\gamma$.
\end{lem}
\begin{proof}
If the $b$-orbit crashes at $x=\bar x$ with rotation
$\theta(x,b)>k\pi$, then there is a $y$ such that $\theta(y,b)=k\pi$.
Let $B=(f(y,b)+1)^2$. By Proposition~7.3, if $a$ is sufficiently
close to $b$, $A {f'}^2$ (along the $a$-orbit) is bounded in Q4 by
$D=\max(\alpha^2/3, B)$; $D$ is a uniform bound on  $A {f'}^2$ in Q4
for all $a$ sufficiently near $b$. Next, choose $x_1$ such that
$0.99\: \bar x < x_1<\bar x$ and such that
$A(x_1,b)<0.01, g(x_1,b)=2\tau>0$, and $h(x_1,b)>0.9$;
this is possible by Lemma~7.4 and
Proposition~7.1. Then, for $a$ sufficiently near $b$ we have
$A(x_1,a)<0.02, g(x_1,a)>\tau>0, f(x_1,a)<f(x_1,b)+0.01/\bar x$ and
$h(x_1)>3/4$. We shall find a $\gamma\in (0,0.01\: \bar x)$ that
works for all $a$, that is, it satisfies $h(x,a)>1/4$ for all $a$
sufficiently near $b$ and $\bar x<x<\bar x +\gamma$. So let $a$
satisfy: i) $A {f'}^2$ (along the $a$-orbit) is bounded by $D$, ii)
$A(x_1,a)<0.02$, iii) $h(x_1,a)>3/4$, and iv) $g(x_1,a)>\tau>0$. If
$h(x,a)>1/4$ for all $x<1.01\: \bar x$ and all $a$ near $b$ we are
done -- let $\gamma=0.01\: \bar x$. Otherwise, we define
$x_2=x_2(a)$, etc. by $h(x_2)=3/4, h(x_3)=1/2, h(x_4)=1/4$, where
$x_2, x_3$, and $x_4$ are the largest values of $x<1.01\: \bar x$
with that property. For $x>x_2$ we have from (2.14a)
$x A f''= x g f- (1+A -h^2) f' \geq -(1+A-h^2) f' \geq -f'/4$ since
$h\leq 3/4$ so $f'' \geq -{f'}^3/(4 x A {f'}^2) \geq -{f'}^3/(4*1.01\:
\bar x D)$ or
$f''/{f'}^2 \geq -f'/(4.04 \:\bar x\: D)$. We now integrate the above
from $x_2$ to $x>x_3$ to get
\begin{equation}
\frac{-1}{f'(x)} \geq \frac{-1}{f'(x)}+\frac{1}{f'(x_2)} \geq
\int_{x_2}^x \frac{f''}{{f'}^2} dx \geq \int_{x_2}^x
\frac{-f'}{4.04\: \bar x\: D} dx = \frac{f(x)-f(x_2)}{4.04\: \bar x\:D}.
\end{equation}
Now, $f(x) \geq f(x_3)$, so $-f'(x) \leq \tfrac{5 \bar x
D}{f(x_2)-f(x_3)} \approx  \tfrac{5 \bar x^2 D}{h(x_2)-h(x_3)}=20\:
\bar x^2 D$. Using the uniform bound on $f'$ in the interval $x_3
\leq x \leq x_4$, we have $x_4-x_3=(f(x_4)-f(x_3))/f'(\xi)$ for some
$\xi\in [x_3,x_4]$. But $(f(x_4)-f(x_3))/f'(\xi) \geq
(h(x_4)-h(x_3))/(\bar x f'(\xi)) \geq 1/80 \:\bar x^3 D$ and hence we
may take $\gamma=1/80\: \bar x^3 D$.
\end{proof}
\begin{lem}
In the interval $x_1<x<\bar x+\gamma$, $g(x,a)>\min(\tau,0.9/h^2(\bar
x,b))$.
\end{lem}
\begin{proof} From (2.18) we have $x g'=C^2 (1-A+x^2 A {f'}^2 - x^2 g
f^2) \geq C^2 (1-A -x^2 g f^2)$. Moreover, since $A(x_1,a)<0.02$ and
$x A'<1, A(x,a)=A(x_1,a)+A'(z) (x-x_1)<0.02 +1/z (0.02 \bar x) <
0.04$, so if $g<0.96/h^2(x)$ then $g'>0$. Since
$f(x_1,a)<f(x_1,b)+0.01/\bar x, h(x,a)\leq 1.01 \:\bar x f(x_1,a) <
1.01 (\bar x f(x_1,b) +0.01) <1.02\: \bar x f(x_1,b)$, we have $g'>0$
if $g(x_1,a)<0.9/h^2(\bar x,b)$. Thus, if $\tau<g(x_1,a)<0.9/h(\bar
x,b), g'>0$, and $g(x,a)>\tau$ in the interval $x_1<x<\bar x
+\gamma$; if $g(x_1,a)>0.9/h^2(\bar x,b)$, then $g(x,a)>0.9/h^2(\bar
x,b)$ for all $x$ in the interval $x_1<x <\bar x +\gamma$ because $g$
cannot cross that value from above.
\end{proof}
Note that the above lower bound on $g$ is uniform -- it applies to
all $a$ satisfying the conditions
 i) $A {f'}^2$ (along the $a$-orbit) is bounded by $D$, ii)
$A(x_1,a)<0.02$, iii) $h(x_1,a)>3/4$, and iv) $g(x_1,a)>\tau>0$.
\begin{lem}
For all $a$ sufficiently near $b$, $v(x,a)$ goes to $0$ for some
$x<\bar x +\gamma$.
\end{lem}
\begin{proof} To show that $v(x,a)$ goes to $0$, we note that $h(x,a)
\geq 1/4$ for all $a$ near $b$ and $ \bar x<x<\bar x +\gamma$ by
Lemma~7.5. Hence, $f(x,a)=h(x,a)/x>1/4 \bar x$. By Lemma~7.6,
$g(x,a)>\min(\tau,1/h(\bar x)$, hence $v'\geq 1/4 \bar x
\min(\tau,1/h(\bar x) =\eta>0$ for $\bar x<x< \bar x +\gamma$. Thus,
$v(\bar x+\gamma)-v(\bar x)= \int_{\bar x}^{\bar x +\gamma} v' dx
\geq \int_{\bar x}^{\bar x +\gamma} \eta dx \geq \eta \gamma$. Let
$x_1$ be chosen so that $v(x_1,a)>-\eta \gamma/2$.  Then, if $a$ is
sufficiently close to $b$, $|v(x_1,a)-v(x_1,b)|>\eta \gamma/2$ so
$v(x_1,a)>-\eta \gamma$. For such $a$ we then have $v(\bar x
+\gamma,a)>v(\bar x,a)+\eta \gamma$ and $v(\bar x,a)>v(x_1,a)>-\eta
\gamma$ because $v'>f g>0$; thus, $v(\bar x +\gamma,a)>0$.
\end{proof}
We now complete the proof of Proposition~7.2.
\begin{proof}[Proof of Proposition~7.2 b)] Suppose that the $b$-orbit
crashes at $x=\bar x$ with $\theta(x,b)<(k+1/2)\pi$ for all $x<\bar
x$ and $\theta(x,b)>k \pi$ for $x$ near $\bar x$. For $a$ near $b$
there is an $x<\bar x +\gamma$ with $v(x,a)=0$ by Lemma~7.7. Since
$x<\bar x +\gamma$, $h(x)>1/4$, i.e., $f(x,a)>0$, so the $a$-orbit
crashes, $A(x,a)=0$, or exits Q4 to Q1 (or Q2 to Q3), $f'(x,a)=0$,
never to return. In either case, the $a$-orbit has rotation
$\theta(x,a)<(k+1/2)\pi$.
\end{proof}
\section{Proof of the main theorem}
\begin{proof}[Proof of Theorem~2.1] Let
$X_n=\{a>0 \:|\: \theta(x,a) <(n+1/2) \pi$ for all $x$\ for which
the a-orbit is defined\}. Note that $X_{n-1}\subset X_n$ and
$X_0\neq \emptyset$ by Proposition~5.1 and hence, $X_n \neq
\emptyset$. Also note that $b_{n+1}>0$ is a lower bound for $X_n$
by Proposition~4.1; hence, $X_n$ has a greatest lower bound
$a_n=\inf(X_n)\geq b_{n+1}>0$. We will show that the $a_n$-orbit
is a globally regular solution and $n\pi<\theta(x,a_n)<(n+1/2)\pi$
for large $x$.

We first show that $a_n\in X_n$, i.e., $a_n$ is the smallest element
in $X_n$. If $\theta(x,a_n)>(n+1/2)\pi$ for some $x$ then
$\theta(x,a)>(n+1/2)\pi$ for all $a$ near $a_n$ so $a\notin X_n$ for
these $a$'s and this contradicts the fact that $a_n$ is the greatest
lower bound of $X_n$. Thus, $a_n\in X_n$. In particular, the
$a_n$-orbit has bounded rotation.

 Next we show that the $a_n$-orbit
does not crash. Recall from Proposition~7.1 that if the $a_n$-orbit
crashes at $x=\bar x$ then $g(x,a_n)>0$ for $x$ near $\bar x$. If the
$a_n$-orbit crashes in Q1 or Q3, that is, if $\theta(x,a_n)<n \pi$
for $x$ near $\bar x$ then $\theta(x,a)<n\pi$ and $g(x,a)>0$ for all
$a$ near $a_n$ which implies by Lemma~2.6b that the $a$-orbit must
have $\theta(x,a)<n\pi$ for all $x$. Thus, $a\in X_n$ for all $a$
near $a_n$ and this contradicts the fact that $a_n$ is the greatest
lower bound of $X_n$.

Similarly, if the $a_n$-orbit crashes in Q2 or Q4, that is, at some
$\bar x$ with $(n+1/2)\pi>\theta(\bar x,a_n)>n \pi$, then by the
crash lemma $(n+1/2)\pi>\theta(x,a)$ for all $x$ in the domain of
definition of the $a$-orbit for all $a$ near $a_n$ and this
contradicts the fact that $a_n$ is the greatest lower bound of $X_n$.

Thus, the $a_n$-orbit is defined for all $x$ and hence is a globally
regular solution by Propositions~6.1. Also, by Proposition~6.2, the
$a_n$-orbit is in Q2 or Q4 for large $x$. It remains to prove that
$\theta(x,a_n)>n \pi$ for large $x$. Suppose that
$\theta(x,a_n)<n\pi$ for large $x$. By Lemma~6.3 we have that
$g(x,a_n)>0$ for large $x$ and hence, $g(x,a)>0$ for all $a$ near
$a_n$. Then, by Lemma~2.6b the $a$-orbit must have $\theta(x,a)<n\pi$
for all $x$ and thus $a\in X_n$, and this contradicts the fact that
$a_n$ is the greatest lower bound for $X_n$. This completes the proof
of Theorem~2.1.
\end{proof}

\section*{Acknowledgment} We would like to thank  the Mathematisches Forschungsinstitut in
Oberwolfach for supporting this project under of the  Research in
Pairs program. P.B. was supported in part by the
KBN~grant~2~P03B~010~16.

\end{document}